\documentclass[aps,amsmath,amssymb,superscriptaddress, twocolumn]{revtex4}
\usepackage{graphicx}
\usepackage{dcolumn}
\usepackage{bm}

\usepackage{colordvi}
\usepackage{color}
\usepackage{ulem}

\newcommand{\TK}{\color{blue}}

\begin{document}

\title{Spin dephasing and  photoinduced spin diffusion  in high-mobility 110-grown GaAs-AlGaAs two-dimensional electron systems}
\author{R.\ V\"olkl}
\affiliation{Institut f\"ur Experimentelle und Angewandte Physik,
Universit\"at Regensburg, D-93040 Regensburg, Germany}
\author{M. Griesbeck}
\affiliation{Institut f\"ur Experimentelle und Angewandte Physik,
Universit\"at Regensburg, D-93040 Regensburg, Germany}
\author{S. A. \ Tarasenko}
\affiliation{A. F. Ioffe Physical-Technical Institute, Russian Academy of Sciences, 194021 St. Petersburg, Russia}
\author{D.\ Schuh}
\affiliation{Institut f\"ur Experimentelle und Angewandte Physik,
Universit\"at Regensburg, D-93040 Regensburg, Germany}
\author{W.\ Wegscheider}
\affiliation{Solid State Physics Laboratory, ETH Zurich, 8093 Zurich, Switzerland}
\author{C.\ Sch\"uller}
\affiliation{Institut f\"ur Experimentelle und Angewandte Physik,
Universit\"at Regensburg, D-93040 Regensburg, Germany}
\author{T.\ Korn}
\email{tobias.korn@physik.uni-regensburg.de}
\affiliation{Institut
f\"ur Experimentelle und Angewandte Physik, Universit\"at
Regensburg, D-93040 Regensburg, Germany}
\date{\today}

\begin{abstract}
We have studied  spin dephasing and spin diffusion in a high-mobility two-dimensional electron system, embedded in a GaAs/AlGaAs quantum well grown in the  [110] direction, by a two-beam Hanle experiment. For very low excitation density, we observe spin lifetimes of more than 16~ns, which rapidly decrease as the pump intensity is increased. Two mechanisms contribute to this decrease: the optical excitation produces holes,  which lead to a decay of electron spin via the Bir-Aranov-Pikus mechanism and recombination with spin-polarized electrons. By scanning the distance between the pump and probe beams, we observe the diffusion of spin-polarized electrons over more than 20~microns. For high pump intensity, the  spin polarization in a distance of several microns from the pump beam is larger than at the pump spot, due to the reduced influence of photogenerated holes.
\end{abstract}

\maketitle

Semiconductor spintronics is a research field which aims at utilizing the electron spin degree of freedom, instead of its charge, for information processing~\cite{Fabian04,WuReview}.
 A key ingredient for  spintronics devices is a semiconductor structure, which allows for long-range  transport of spin-polarized carriers as well as manipulation of the spin polarization via the Rashba spin-orbit interaction~\cite{Rashbafield}. While several studies on spin injection and spin transport in GaAs have focussed on weakly n-doped GaAs bulk layers~\cite{awschalom99,crooker07,PhysRevB.79.245207}, the mobility in these structures is very low due to the direct doping, and the bulk-like character of the layer does not allow for direct spin modulation via the Rashba effect. By contrast, in two-dimensional electron systems (2DES) based on modulation doping of quantum wells (QWs), very high mobilities can be reached, and electrical control of the spin dynamics via the Rashba effect has been demonstrated~\cite{Harley03}. Structures grown along the [110] crystallographic direction are particularly promising for spintronics applications, as the Dresselhaus spin-orbit field points along the growth direction and spin dephasing via the Dyakonov-Perel (DP) mechanism~\cite{DP} is strongly suppressed~\cite{DP110,Ohno99_1,Oestreich04_1}, or modified, if there is also Rashba spin-orbit interaction due to growth-axis asymmetry~\cite{tarasenko:165317}.

Here, we present spatially resolved studies of the spin dynamics of optically oriented electrons in a high-mobility 2DES embedded in a (110)-grown  GaAs/AlGaAs QW structure. Our sample is a 30~nm wide QW  with  electron density  $n_e= 2.4\times 10^{11}$~cm$^{-2}$ and mobility $\mu = 2.3\times 10^{6}$~cm$^{2}$/Vs at 1.5~K, determined by magnetotransport measurements. The modulation doping is placed symmetrically above and below the QW in a complex structure similar to that used by Umansky et al.~\cite{Umansky20091658}.  Similar growth procedure leads to even higher carrier mobilities in (001)-oriented QWs and allows for the study of ballistic cyclotron motion~\cite{Griesbeck09}. Due to the low growth temperature for the (110) surface, symmetrical remote doping results in a symmetrical band profile of the QW  and vanishing Rashba interaction~\cite{bel'kov:176806,olbrich:245329}.  The sample is mounted in vacuum on the cold finger of a He-flow cryostat, the electron temperature in the sample during measurements has been determined by analyzing the high-energy tail of the photoluminescence (see, e.g., \cite{KornReview} for details) to be about 20~K .  A cw diode laser is used as a source for the circularly polarized pump beam. It  nonresonantly excites spin-polarized electron-hole pairs in the QW  with an excess electron energy about 25~meV above the Fermi energy.  The $z \parallel [110]$ component of the spin polarization is detected by near-resonant probing with a tunable Ti-Sapphire laser, which is linearly polarized, via the magneto-optic Kerr effect (MOKE). Both beams are superimposed on each other at a beamsplitter and either coupled into a 20x microscope objective or focussed onto the sample using a lens with 50~mm focal length. The reflected probe beam is spectrally filtered using a bandpass to suppress the collinear pump beam and then coupled into an optical bridge, which detects the small rotation of the probe beam polarization axis due to the polar MOKE. A lock-in modulation scheme is used to increase the sensitivity of the detection. The spot sizes of pump and probe beams are 6~$\mu$m and 3~$\mu$m, respectively, if the microscope objective is used. The lens leads to larger spot sizes of 40~$\mu$m for both beams.  The spot size is determined by scanning the beam over a lithographically defined structure. For spatially resolved measurements, the pump beam is scanned with respect to the probe beam by a piezo-controlled mirror. The cryostat is mounted between a pair of Helmholtz coils, and magnetic fields up to 30~mT can be applied in the sample plane.

First, we  discuss the electron spin dephasing at the overlap of pump and probe spots as a function of the pump intensity.
\begin{figure}
  \includegraphics[width= 0.35\textwidth]{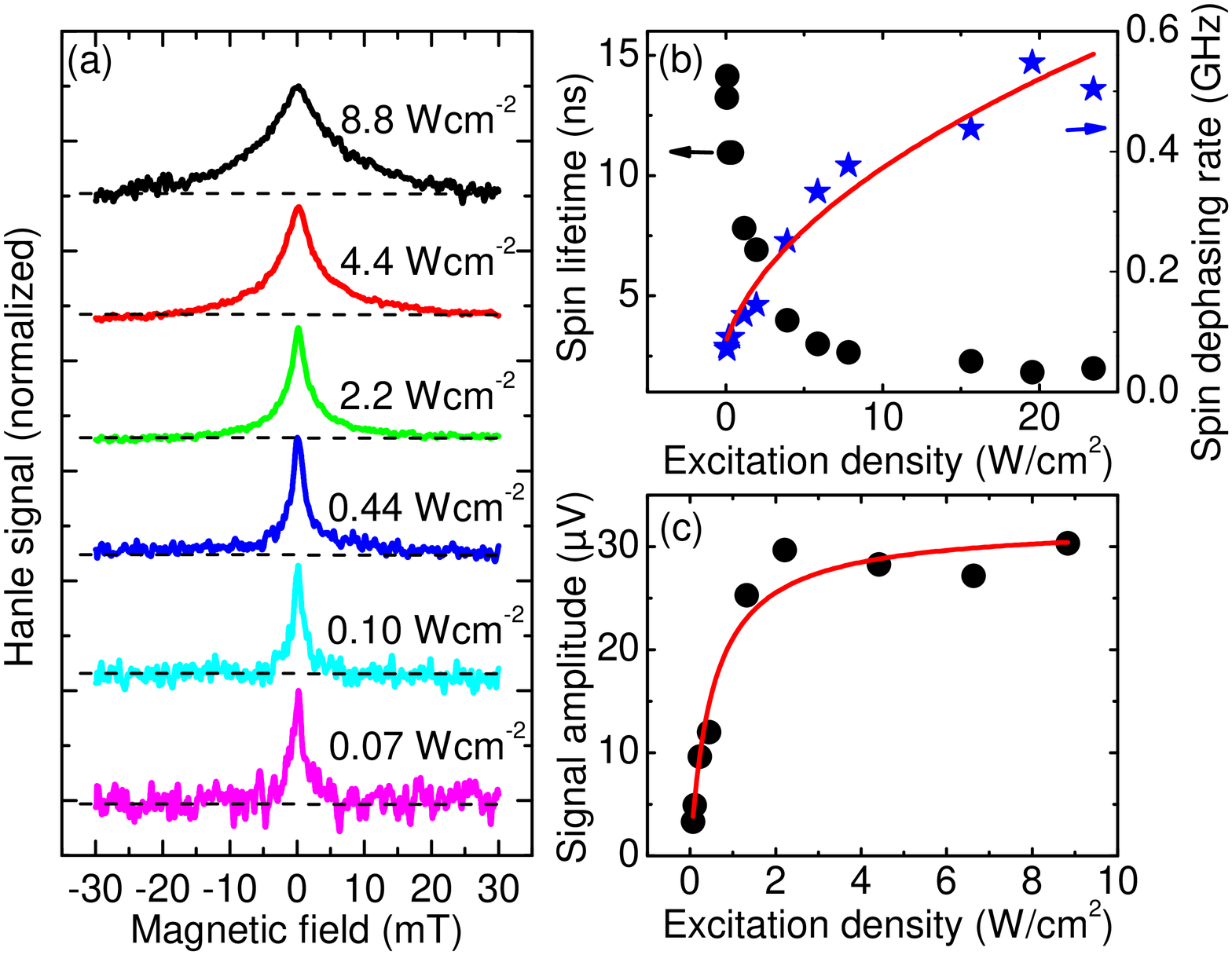}
   \caption{(color online) (a) Hanle-MOKE traces measured  for different excitation densities.(b) Spin lifetime $\tau_s$ (dots) and spin dephasing rate (stars) as a function of excitation density. The solid red line is a fit to  Eq.~\ref{tau_s_Hanle5} with $I_0 = 0.53$~W/cm$^2$ . (c) Hanle-MOKE signal amplitude as a function of the excitation density. The solid red line is  plotted after  Eq.~\ref{S_zerofield3} with the same $I_0$.}
   \label{Power_Curves}
\end{figure}
Figure~\ref{Power_Curves}(a) shows a series of Hanle-MOKE traces measured  for different excitation densities of the pump beam. The traces have been normalized for easier comparison. All traces show the typical Lorentzian lineshape of a Hanle curve: at zero magnetic field,  the $z$ component of the total spin $\bm{S}$
 has a maximum. An applied in-plane magnetic field leads to spin precession and decrease of $S_z$. The spin lifetime $\tau_s$ can be extracted from the half-width of the Lorentzian, which is described by
\begin{equation}\label{Hanle_B}
S_z(B) = \frac{S_z(0)}{1+ (\omega_L \tau_s)^2}.
\end{equation}
Here, $\omega_L=g_e \mu_B B/\hbar$ is the Larmor frequency, which is determined by the electron $g$ factor and the applied magnetic field $B$, and $\mu_B = e \hbar/2m$ is the Bohr magneton. The  in-plane  $g$ factor of our sample has been determined by time-resolved Faraday rotation measurements (not shown) to be $|g_e|=0.38$. The series of Hanle-MOKE traces clearly shows how the Lorentzian widens with increasing the excitation density of the pump beam. Figure~\ref{Power_Curves}(b) shows the  extracted spin lifetime and the corresponding dephasing rate, {\TK $1/\tau_s$}, as a  function of the excitation density. For the lowest excitation density used, we  find a spin lifetime of 14.1~ns, which is  of the same order as values recently observed for (110)-grown QWs by spin noise spectroscopy~\cite{oestreich08,Müller2010569} in small in-plane magnetic fields. As the excitation density is increased, the spin lifetime drops significantly, and the dephasing rate increases accordingly.  A similar behavior of the spin lifetime as a function of excitation density is observed in measurements  with  a large focal spot of 40~$\mu$m (not shown). The larger total laser power used in these measurements allows us to reach even lower excitation densities, resulting in a maximum observed  $\tau_s$  of 16.4~ns.

We also  analyze the amplitudes of the Hanle curves, which provide information on the spin polarization at zero magnetic field. It is observed that the amplitude, depicted in Figure~\ref{Power_Curves}(c), saturates for   excitation densities above 1~Wcm$^{-2}$.

In order to interpret  both  observations,  the  decrease of $\tau_s$ and the saturation of $S_z(0)$ with the excitation density, we consider the spin dynamics of optically oriented electrons.
For simplicity, we neglect the influence of inhomogeneous pumping and (spin) diffusion on the  spin dynamics. This is justified, as our measurements using a large focal diameter spot yield similar values for the spin dephasing rates and show the same functional dependence on intensity as the measurements using the smaller focal diameter.

At zero magnetic field,  the  $z$ component of the steady-state total electron spin  is given by
\begin{equation}\label{S_Z}
S_z(0) = G_z \tau_z \:,
\end{equation}
 where  $G_z$ is the spin generation rate, proportional to the excitation  density $I$, and $\tau_z$ is the lifetime of electron spin oriented along the QW normal. The decay rate $1/\tau_z$  is determined by three  contributions,
\begin{equation}\label{tau_sz}
1/\tau_{z} = 1/\tau_{z}^{lim} + \gamma_z^{BAP} N_h + \gamma^{r} N_h \:,
\end{equation}
with $1/\tau_{z}^{lim}$  being  the spin dephasing rate in the limit of zero excitation. In this regime, spin dephasing may occur either due to the Elliott-Yafet mechanism (EY)~\cite{Yafet63} or due to the DP mechanism caused by small random Rashba  fields in a modulation-doped structure~\cite{Sherman_Rashba, Wu_randomRashba}. The terms
$\gamma_z^{BAP} N_h$ and $\gamma^{r} N_h$ describe the spin decay due to the Bir-Aronov-Pikus (BAP) mechanism~\cite{BAP}, which is proportional to the steady-state hole density $N_h$, and the recombination of optically oriented electrons with holes, respectively. The role of photocarrier recombination is as follows: in the limit of high excitation density, nearly all of the optically oriented electrons  would recombine with holes, similar to an undoped QW, and the photocarrier lifetime limits the spin lifetime. It is assumed here that the recombination probability for an electron  is also proportional to the steady-state hole density, which is usually valid for a non-degenerate hole gas in the absence of recombination centers,  and $N_h \ll n_e$. We also note that, for simplicity, we neglect the influence of photoexcitation on the EY and DP mechanisms.

It follows  from Eqs.~(\ref{S_Z}) and~(\ref{tau_sz}) that $S_z(0)$ is given by
\begin{equation}\label{S_zerofield2}
S_z(0) = \frac{G_z}{1/\tau_{z}^{lim} + (\gamma_z^{BAP} + \gamma^{r}) N_h}  \:.
\end{equation}
In the limit of very low excitation density, where  the time $\tau_{z}^{lim}$ determines the spin lifetime,  $S_z (0)$ depends linearly on the excitation density $I$, as $G_z \propto I$. For increased excitation density, $S_z$ saturates, as $N_h \propto I$. Due to the low spin dephasing rate $1/\tau_{z}^{lim}$, this saturation already occurs at rather low values of the excitation density, as observed in Figure~\ref{Power_Curves}(c). The experimental amplitude data is well-described by the fit function
\begin{equation}\label{S_zerofield3}
S_z(0) \propto \frac{I}{1 + I/I_0}  \:,
\end{equation}
which has the excitation density dependence of  Eq.~(\ref{S_zerofield2}). We note that, from the excitation-density dependence, the relative magnitude of $\gamma_z^{BAP}$ and $\gamma^{r}$ cannot be determined, but we may conclude that $1/\tau_{z}^{lim}$ and $(\gamma_z^{BAP} + \gamma^{r}) N_h$ are  already comparable at an excitation level of less than $I \sim 1$~W/cm$^2$.
\begin{figure}
  \includegraphics[width= 0.35\textwidth]{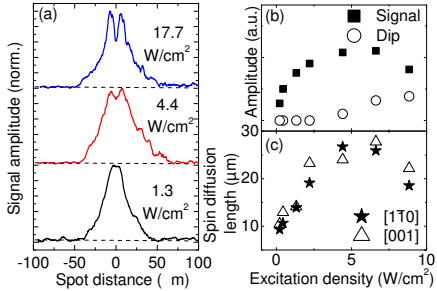}
   \caption{(color online) (a) Scanning Hanle-MOKE measurements for different excitation densities. The curves have been normalized for easy comparison. (b)Amplitude of the Hanle-MOKE signal  at the beam overlap (squares) and the dip amplitude (circles) as functions of the excitation density. (c) Spin diffusion length as a function of excitation density. The different scan directions are indicated by triangles ($[001]$)  and  stars ($[1\bar{1}0]$).}
   \label{Diff_Amp}
\end{figure}

\begin{figure}
  \includegraphics[width= 0.35\textwidth]{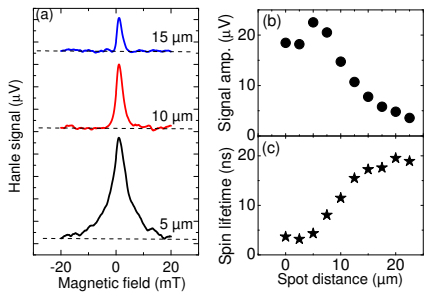}
   \caption{(color online) (a)  Hanle-MOKE traces for different distances between pump and probe beams.  (b)Signal amplitude  of the Hanle-MOKE traces  as a function of pump-probe distance. (c) Spin lifetime extracted from the Hanle-MOKE traces as a function of pump-probe distance.}
   \label{Fullcurves}
\end{figure}

Next, we consider the spin dephasing in an applied in-plane magnetic field. The spin lifetime obtained in Hanle measurements is given by
\begin{equation}\label{tau_s_Hanle}
 \tau_s = \sqrt{\tau_{z} \tau_{\|}} \:,
\end{equation}
where $\tau_{\|}$ is the in-plane spin dephasing time. The rate $1/\tau_{\|}$ can be also presented in the form of Eq.~(\ref{tau_sz}).
However, $1/\tau_{\|}^{lim} \gg 1/\tau_{z}^{lim}$   in (110)-grown structures, as $1/\tau_{\|}^{lim}$ is determined by the conventional DP mechanism in the Dresselhaus field perpendicular to the  QW plane, while $\gamma_{\|}^{BAP}$ and $\gamma_{z}^{BAP}$ are comparable. For symmetrically (110)-grown structures, the experimentally observed ratio $\tau_z^{lim} / \tau_{\|}^{lim} $ is about 1 order of magnitude~\cite{oestreich08}.  In the excitation density range we study, we may therefore neglect the effects of BAP mechanism and recombination on the in-plane spin dephasing rate, and use the following expression for the spin lifetime measured in Hanle experiments
\begin{equation}\label{tau_s_Hanle4}
1/\tau_{s} = \sqrt{1/\tau_{\|}^{lim}} \sqrt{[1/\tau_{z}^{lim} + (\gamma_z^{BAP} + \gamma^{r}) N_h]} \:.
\end{equation}
For $N_h \propto I$, the  spin dephasing rate extracted from the Hanle  curves  has the following intensity dependence
\begin{equation}\label{tau_s_Hanle5}
1/\tau_{s}(I) = 1/\tau_{s}(0) \sqrt{ 1 + I/I_0} \:
\end{equation}
with the same intensity $I_0$ as in Eq.~(\ref{S_zerofield3}).
As Fig.~\ref{Power_Curves}(b) shows, the measured rates are in good agreement with  Eq.~(\ref{tau_s_Hanle5}). Both, the intensity dependence of the spin dephasing rate, and the signal amplitude [Fig.~\ref{Power_Curves}(c)], are well described by fits using the value $I_0= 0.53$~Wcm$^{-2}$.

Finally, we investigate the spin diffusion in our sample. Fig.~\ref{Diff_Amp}~(a) shows a series of scanning Hanle-MOKE measurements. The traces were generated in the following way: at each distance between pump and probe spot, the difference between the Kerr signals for zero field and 30~mT in-plane fields was recorded, and this amplitude, which is proportional to the  local spin polarization, is plotted as a function of the spot distance.  For low excitation density, we observe a maximum of the spin polarization at the beam overlap. The spin polarization decays monotonously as the distance between pump and probe is increased. For higher excitation densities, however, a local minimum becomes visible at the beam overlap, which becomes more pronounced as the excitation density is increased. We define the amplitude of this dip as the signal difference between the local minimum and the adjacent absolute maxima at finite pump-probe beam distance.   Spin polarization at the beam overlap saturates below  3~W/cm$^2$, above this value the dip develops and increases its amplitude, as Fig.~\ref{Diff_Amp}~(b) demonstrates.

In order to understand this anomalous  spin distribution, we consider the strongly different diffusion coefficients of electrons and heavy holes in a GaAs-based QW structure: the lighter spin-polarized electrons  rapidly diffuse away from the pump spot due to their low mass, while the heavy holes move more slowly. The  electron diffusion is significantly faster in this high-mobility 2DES as compared to, e.g.,  bulk n-GaAs, due to the low concentration of momentum scattering sites and the large Fermi wave vector of the electrons. Consequently, in a distance of a few microns from the pump spot, which is smaller then the diffusion length of spin-polarized electrons, the hole density is significantly reduced. Therefore,  hole-dependent spin dephasing and recombination processes are suppressed, resulting in a larger  electron spin polarization. We note that the width of the observed dip increases only slightly from about 8~$\mu$m to 11~$\mu$m as a function of the excitation density, indicating the slow diffusion of the optically created holes. The spin diffusion length, $l_s$,  ( determined by fitting an exponential decay of the spin polarization with distance far from the beam overlap, so that the signal dip may be neglected) increases first with the excitation density, as Fig.~\ref{Diff_Amp}~(c) shows, reaching a maximum of about 27~$\mu$m, then decreases again for higher excitation densities. The two in-plane crystallographic directions $[001]$  $[1\bar{1}0]$ show a comparable behavior. Several effects are likely to contribute to the initial increase. First, the inhomogeneous distribution of electrons and holes in the QW plane produces local electric fields. Far away from the pump spot, the field is directed towards the spot and induces a divergent electron flow enlarging the area of spin polarization. Second, with the local increase of the electron density, high-$k$ states are occupied by the optically oriented electrons.  Additionally, the rise of spin polarization suppresses electron-electron collisions between particles of opposite spins, which slow down the spin diffusion~\cite{Damico00,Weber05}, thereby increasing $l_s$. For very high excitation densities,  the spin diffusion length is decreased due to local heating of the sample and electron-electron scattering.
To study the spatial dependence of the spin lifetime in more detail, we perform Hanle-MOKE measurements for high excitation density (5~Wcm$^{-2}$) as a function of the pump and probe beam distance. Fig.~\ref{Fullcurves}(a) shows a series of Hanle-MOKE traces for different distances between pump and probe. We clearly observe that the Hanle curves become more narrow as the distance increases. This corresponds to an increase of the  effective  spin lifetime from below 4~ns at the beam overlap to a saturation value of about 18~ns, which is reached at a distance of 15~$\mu$m between pump and probe (Fig.~\ref{Fullcurves}(c)). A spatially inhomogeneous spin lifetime due to local optical pumping was previously observed in  bulk n-GaAs~\cite{PhysRevB.79.245207}.   The amplitude of the Hanle signal extracted from these traces, shown in Fig.~\ref{Fullcurves}(b),  mimics the scanning Hanle-MOKE measurements for high excitation density (Fig.~\ref{Diff_Amp}(a)), in that the signal amplitude has a dip at the beam overlap and then first increases at a distance of 5~$\mu$m, then decreases again.

In conclusion, we have investigated the spin dephasing and spin diffusion in a high-mobility (110)-grown two-dimensional electron system by spatially resolved Hanle-MOKE measurements. We observe long spin dephasing times at low excitation density, which are strongly reduced for higher excitation. The spatially resolved measurements show very long spin diffusion length  up to 27~$\mu$m. For high excitation density, the  spin polarization is larger at some distance from the pump laser spot than at the overlap of pump and probe beams due to the reduced spin dephasing in the absence of optically created holes.
The authors would like to thank E.L. Ivchenko, M.M. Glazov, J.-H. Quast, T. Kiessling, W. Ossau and E.Ya. Sherman for fruitful discussion.  Financial support by the DFG via SPP 1285 and SFB 689 , RFBR, and ``Dynasty'' Foundation is gratefully acknowledged.
\bibliography{Diff110}
\end{document}